\documentclass[AMS,STIX1COL]{WileyNJD-v2}
\usepackage{moreverb}
\usepackage[version=4]{mhchem}

\newcommand\BibTeX{{\rmfamily B\kern-.05em \textsc{i\kern-.025em b}\kern-.08em
T\kern-.1667em\lower.7ex\hbox{E}\kern-.125emX}}

\articletype{Letters}%

\received{<day> <Month>, <year>}
\revised{<day> <Month>, <year>}
\accepted{<day> <Month>, <year>}

\begin{document}

\title{Thermal lensing-induced soliton molecules in $\beta$-phase Gallium Oxide}

\author[1]{Alain M. Dikand\'e*}

\author[1]{E. Aban Chenui}

\author[2]{A. Sunda-Meya}

%\authormark{AUTHOR ONE \textsc{et al}}

\address[1]{Laboratory of Research on Advanced Materials and Nonlinear Science (LaRAMaNS),
Department of Physics, Faculty of Science, University of Buea P.O. Box 63 Buea, Cameroon.}

\address[2]{Nanoscale Engineering Lab (NanoEL), Department of Physics and Computer Science, Xavier University of Louisiana, 1 Drexel Drive, New Orleans, Louisiana 70125, USA.}

\corres{*Alain Mo\"ise Dikand\'e, Corresponding author. \email{dikande.alain@ubuea.cm}}

%\presentaddress{Present address}

\abstract[Abstract]{In recent years, beta gallium oxide ($\beta$-\ce{Ga2O3}) has become the most investigated isomorph of gallium oxide polymorphs, due to the great potential it represents for applications in optoelectronics and photonics for solar technology, particularly in blind ultraviolet photodetector solar cells (SBUV) designs. To optimize its use in these applications, and to identify possible new features, knowledge of its fundamental properties is relevant. In this respect, optical, thermal and electronic properties of $\beta$-\ce{Ga2O3} have been studied expriementally, providing evidence of a wide-band inorganic and transparent semiconductor with a Kerr nonlinearity. Thermo-optical properties of the material, probed in SBUV sensing experiments, have highlighted a sizable heat diffusion characterized by a temperature gradient along the path of optical beams, quadratic in beam position and promoting a refractive-index change with temperature. The experimentally observed Kerr nonlinearity together with the thermally induced birefringence, point unambiguously to a possible formation of soliton molecules during propagation of high-intensity fields in $\beta$-\ce{Ga2O3}. To put this conjecture on a firm ground we propose a theoretical analysis, based on the cubic nonlinear Schr\"odinger equation in 1+1 spatial dimension, in which thermal lensing creates an effective potential quadratic in the coordinate of beam position. Using the non-isospectral inverse-scattering transform method, the exact one-soliton solution to the propagation equation is obtained. This solution features a bound state of entangled pulses forming a soliton molecule, in which pulses are more or less entangled depending on characteristic parameters of the system.}


\jnlcitation{\cname{%
\author{Chenui Aban E.}, \author{Dikand\'e Alain M.}, and \author{A. Sunda Meya}} (\cyear{2021}), 
\ctitle{Thermal lensing-induced soliton molecules in $\beta$-phase Gallium Oxide}, \cjournal{Microw Opt Technol Lett.}, \cvol{2021;00:0--0}.}

\maketitle

%\footnotetext{\textbf{Abbreviations:} ANA, anti-nuclear antibodies; APC, antigen-presenting cells; IRF, interferon regulatory factor}

\section{Introduction}
\label{intro}
Gallium oxide (\ce{Ga2O3}) belongs to a family of emergent materials with exotic structural, electronic and thermal properties that attracted a great deal of attention over the past two decades \cite{r4a,r4,r4b,r5,r6,r7,r8,r9,r10}. Due to their outstanding properties, these materials find widespread applications in modern telecommunication technology including optoelectronics, nanotechnology, photonic sensors and nonlinear ultrashort optical pulse processings \cite{r4,r4b,r6,r7}. \ce{Ga2O3} is an inorganic solid that exists in five dictinct water-insoluble crystalline forms which are $\alpha$, $\beta$, $\delta$, $\gamma$ and $\epsilon$ \cite{r4a,r4b,r5}. Of these five polymorths $\beta$-\ce{Ga2O3} \cite{r11,r11a} is the most stable crystalline form, with a melting point of 1900 $^\circ$C. In $\beta$-\ce{Ga2O3}, oxide ions are located in a distorted cubic closest packing arrangement, while Gallium ions occupy distorted tetrahedral and octahedral sites with Ga-O bond distances of $1.83 \AA$ and $2.0 \AA$, respectively \cite{r4a}. This closest packing arrangement confers to $\beta$-\ce{Ga2O3} a crystal structure characteristic of inorganic close-packed solids with strongly anisotropic structures \cite{r4a,r11a,r12}. 
\par Optical properties of $\beta$-\ce{Ga2O3} have been characterized in several recent experiments \cite{r4b,r12aa,r12a,r12b,r12c,r12d,r12e} and it is known that this material is a transparent semiconducting oxide. Experimental measurements of two-photon absorption coefficient and Kerr nonlinear coefficient, as well as their polarization dependence along (010), have been carried out \cite{r12a} and the wavelength dependence of two-photon absorption coefficient together with the Kerr nonlinear refractive index were estimated \cite{r12a}. Results suggest that $\beta$-\ce{Ga2O3} exhibits a two-photon absorption coefficient of 1.2 cm/GW, and a Kerr nonlinear refractive index of $n_K=2.1 \times 10^{-15} cm^2/W$ along (010). Moreover $\beta$-\ce{Ga2O3} shows stronger in-plane nonlinear optical anisotropy on (201) plane than on (010) plane \cite{r12a}, suggesting that $\beta$-\ce{Ga2O3} has the potential for ultra-low loss waveguides and ultra-stable resonators and integrated photonics, especially in UV and visible wavelength spectral range \cite{r12a,r13,r14}.\par
Another aspect of great interest in recent investigations of optical properties of $\beta$-\ce{Ga2O3}, is their intricate connections with heat (i.e. thermal) transport in the material \cite{r11a,r12c,r13,r15}. In this respect it was found that due to the quasi-one-dimensional anisotropy, thermal conductivity in $\beta$-\ce{Ga2O3} is the highest along (010) at all temperatures used in measurements \cite{r12}. Thus the thermal conductivity, measured by means of laser-
flash methods, gives 21 W/mK along the (010) direction \cite{r12}. Photo-thermal effects induced by optical beams propagating along the direction of main thermal conductivity of the material, discussed in ref. \cite{r12c} both experimentally and theoretically, have established a linear dependence of the effective refractive index on temperature. Based on their experiments the authors postulated that the correction to the refractive index, due to thermo-optical effect, could be expressed \cite{r12c}:
\begin{equation}
 \Delta n(T)= \tilde{\beta}\, \Delta T, \label{onea}
\end{equation}
reflecting a physical process best known as thermal-lensing effect \cite{r16,r17,r17a}. In the above formula the quantity $\tilde{\beta}$ is the thermo-optic coefficient, and $\Delta T$ is the temperature distribution along the path of the optical beam propagating in the material. Experimental values of the thermo-optic coefficient $\tilde{\beta}$ for $\beta$-\ce{Ga2O3}, were measured in ref. \cite{r12c}.\par  
We focus attention on the context of heat storage in the material by the propagating optical beam, at a constant rate. As pointed out in ref. \cite{r13}, in such phyisical context the spatial distribution of temperature in $\beta$-\ce{Ga2O3} can readily be represented by the second-order ordinary differential equation:
\begin{equation}
\frac{d^2 T}{dx^2}=-\frac{Q}{\kappa b}, \label{two}
\end{equation}
where $Q$ is the homogeneous heat flux induced by beam absorption (here assumed constant), $\kappa =$ 27 W/(m K) is the thermal conductivity of $\beta$-\ce{Ga2O3} crystal \cite{r12}, $b$ is material thickness \cite{r13} and $x$ denotes a position along the beam (i.e. along (010)). By integrating eq. (\ref{two}) with boundary conditions:
\begin{equation}
T (x = 0) = T_0, \qquad \frac{d T}{d x}(x=0) =\gamma, \label{bcond}
\end{equation}
where the real and constant coefficient $\gamma$ denotes the rate of change of temperature along the beam propagation direction, we obtain:
\begin{eqnarray}
\Delta T(x)&=& T_0 - T(x) \nonumber \\
&=& \frac{Q}{2\kappa b}x^2 - \gamma\, x.  \label{3}
\end{eqnarray}
It is worth stressing that because the heat gradient is developed along the path of a polarized optical beam, the quantity $\gamma$ can take positive or negative values reflecting possible heat flow either in the direction or opposite to the transverse polarization of the optical beam.\par
Our objective in this work is to examine what the above thermo-optical properties, translate concretely for optoelectronic applications of $\beta$-\ce{Ga2O3} in integrated nonlinear optics and nonlinear photonics \cite{r12a,r18}. In this purpose we consider the propagation of a high-intensity optical beam in $\beta$-\ce{Ga2O3} crystal, in the presence of thermal-lensing effect. The mathematical model will be the standard nonlinear wave equation, derived from Maxwell's equations for an anisotropic optical medium \cite{kiv1,nlq} with a Kerr refractive index modified by thermo-optical processes. 
\section{Propagation equation and exact one-soliton solution}
Consider an electromagnetic field $\mathcal{E}(x,z,t))$ propagating on the $xy$ plane of a thin $\beta$-\ce{Ga2O3} crystal, assuming that its envelope changes very slowly along $x$. In the paraxial approximation we can pick $\mathcal{E}(x,z,t))=u(x,z)e^{-(kz - i\omega t)}$, where $u(x,z)$ is the optical field envelope, $k$ is the wave number and $\omega$ is the carrier frequency. In the paraxial approximation, the envelope $u(x,z)$ of the optical field propagating in the transparent medium with Kerr nonlinearity, obeys the following nonlinear Helmholtz equation in (1+1) dimensions \cite{nlq}:
\begin{equation}
\frac{\partial^2 u}{\partial x^2}+2ik\frac{\partial u}{\partial z} + \frac{2k^2}{n_0}\left[n_K\vert u\vert^2 + \Delta n(T)\right] \,u=0, \label{hel1}
\end{equation}
where $n_K$ denotes the Kerr nonlinearity coefficient and $n_0$ is the unperturbed refractive index (defined as $n_0= k c/\omega$, with $c$ the speed of light). Replacing $\Delta T(x)$ given by eq. (\ref{3}) in the refractive index correction eq. (\ref{onea}), and substituting the latter quantity in the (1+1) dimensional Helmholtz equation (\ref{hel1}), we obtain:
\begin{equation}
\frac{\partial^2 u}{\partial x^2}+2ik\frac{\partial u}{\partial z} + 2g\left[ \vert u\vert^2 -V(x)\right]u=0, \label{hel2}
\end{equation}
where we defined:
\begin{equation}
 g=\frac{n_K}{n_0}k^2, \label{nu}
 \end{equation}
 and;
\begin{eqnarray}
 V(x)&=&-\frac{\mu}{2}x^2 + \mu_0\, x, \label{coefa} \\
 \mu&=&\frac{\tilde{\beta} Q}{\kappa b\,n_K}, \hskip 0.3truecm \mu_0=\frac{\tilde{\beta} \gamma}{n_K}. \label{coef}
\end{eqnarray}
Equation (\ref{hel2}) is a perturbed cubic nonlinear Schr\"odinger equation, where the perturbation is introduced by the external potential $V(x)$ which depends on the spatial variable $x$. Unlike the homogeneous cubic nonlinear Schr\"odinger equation which is known to be exactly integrable \cite{l2,l1}, in general the cubic nonlinear Schr\"odinger equation with an external potential does not admit exact soliton solutions for arbitrary expressions of $V(x)$. Notably, for the particular form of $V(x)$ given by formula (\ref{coefa}), we can introduce the following new variables:
\begin{eqnarray}
s&=&x\sqrt{g}, \hskip 0.3truecm \tau=\frac{g}{2k}\,z, \nonumber \\
F(s)&=&-\frac{L^2}{2}\,s^2 + L_0\,s, \label{inha} \\
    L^2&=&\frac{\mu}{g}, \hskip 0.3truecm L_0=\frac{\mu_0}{\sqrt{g}}. \label{inh}
\end{eqnarray}
Also we redefine the field envelope using a different parameter i.e. $q(s,\tau) \equiv u(x,z)$, such that the inhomogeneous cubic nonlinear Schr\"odinger equation (\ref{hel2}) becomes:
\begin{equation}
iq_{\tau} + q_{ss} + 2\left[\vert q\vert^2 - F(s)\right]q=0. \label{hel3a}
\end{equation}
Equation (\ref{hel3a}) is closely similar to the inhomogeneous cubic nonlinear Schr\"odinger equation considered by Balakrishnan in ref. \cite{ra}, for which the author proposed a non-isospectral inverse-scattering transform approach for the associated exact one-soliton solution. Instructively the non-isospectral inverse-scattering transform \cite{ra} rests essentially on mapping the inhomogeneous cubic nonlinear Schr\"odinger equation (\ref{hel3a}), onto a pair of linear eigenvalue equations similar to the Lax pair \cite{l2,l1}, for which exact eigenvalues and the associated eigen functions exist. If they indeed exist the Lax-pair eigenvalue problem is said to be exactly integrable, and so is the generic inhomogeneous nonlinear Schr\"odinger equation. Balakrishnan coined out \cite{ra} that because of the particular quadratic dependence of the external potential $F(s)$ in the spatial coordinate $s$, eigenvalues of the Lax pair cannot be constant but will depend on $\tau$. In other words the Lax pair in this particular case is a non-isospectral eigenvalue problem \cite{re,rf}. \par
In the specific case of eq. (\ref{hel3a}), the non-isospectral inverse-scattering transform \cite{ra} assumes an initial beam $q(s, \tau = 0)$ which is intented to also serve as the scattering potential. The appropriate choice of reflectioness potential for the Lax pair associated with eq. (\ref{hel3a}), is the following sech-shaped initial field envelope \cite{ra}: 
\begin{equation}
q(s, 0)= U_0\vartheta(s, 0)\,\text{sech}\left[2\int_0^s{\vartheta(y, \tau=0)\,dy}\right], \label{sol1}
\end{equation}
where $U_0$ is a normalization constant and $\vartheta(s,\tau)$ is the eigenvalue of the spectral problem. As stressed the eigenvalue $\vartheta(s,\tau)$ is not homogeneous, but depends on the variables $s$ and $\tau$ consistently with the non-isospectral feature of the Lax pair associated with the inhomogeneous nonlinear Schr\"odinger equation (\ref{hel3a}). Most generally the eigenvalue $\vartheta(s,\tau)$ will be determined by solving the following nonlinear partial differential equation:
\begin{equation}
\frac{\partial \vartheta}{\partial \tau}=-2\frac{\partial (i\vartheta)^2}{\partial s} + \frac{\partial F(s)}{\partial s}, 
\end{equation}
in ref. \cite{ra} it was shown that the corresponding solution can be expressed as a factor of two functions of different variables, namely:
\begin{equation}
\vartheta(s,\tau)=G(s)H(\tau). \label{fac}
\end{equation}
The two functions of distinct variables $G(s)$ and $H(\tau)$ are solutions to the following independent equations:
\begin{eqnarray}
\frac{\partial G^2}{\partial s}&=& \lambda_1\, G, \hskip 0.3truecm \frac{\partial F}{\partial s}=\lambda_0\,G, \label{eignG} \\
\frac{\partial H}{\partial \tau}&=& 2\lambda_1 H^2 + \lambda_0, \label{eignH}
\end{eqnarray}
with $\lambda_1$ and $\lambda_0$ being two arbitrary (we assume so for now) real constants. The set of two ordinary differential equations (\ref{eignG}) leads to the following solution for $G(s)$:
\begin{equation}
G(s)=\frac{\lambda_1}{2}\,s + G(0),  
\end{equation}
provided with the constraint:
\begin{equation}
F(s)=\frac{\lambda_1\lambda_0}{4}\,s^2 +\lambda_0G(0)\,s. \label{bG}
\end{equation}
Quite remarkably the quantities $\lambda_1$ and $\lambda_0$ in eq. (\ref{eignG}), which we referred to as arbitrary real constants, turn out to be actually non arbitrary, in fact they can be expressed as functions of real physical parameters by equating eq. (\ref{eignG}) and (\ref{inha}). This gives:
\begin{equation}
G(0)=L_0, \hskip 0.3truecm \lambda_1=-2L^2, \hskip 0.3truecm \lambda_0=1, \label{bGa}
\end{equation}
where $L$ and $L_0$ are defined in eq. (\ref{inh}). Having the analytical expression of $G(s)$ let us turn to the one for $H(\tau)$, which is found by solving the ordinary differential equation (\ref{eignH}):
\begin{eqnarray}
H(\tau)&=&H(0) \left[\frac{1+\alpha\tanh\left(\Omega \tau \right)}{1 + \alpha^{-1}\tanh\left(\Omega \tau \right)}\right], \label{solH} \\
\Omega&=& 2L, \hskip 0.3truecm \alpha=\frac{1}{\Omega\, H(0)}. \label{cofsolH}
\end{eqnarray}
Knowing the analytical expression of the eigenvalue $\vartheta(s,\tau)$, to generate the complete solution $q(s, \tau)$ to the inhomogeneous nonlinear Schr\"odinger equation (\ref{hel3a}) we follow a standard procedure. The procedure, detailed in ref. \cite{ra}, consists in reconstructing the complete solution $q(s, \tau)$ from the initial profile (\ref{sol1}) by means of a transform-integral equation, so-called Gelfand-Levitan-Marchenko equations, based on two kernels which are matrices determined by the set of scatttering data \cite{ra}. In our specific context this procedure gives rise to the one-soliton solution:   
\begin{equation}
q(s, \tau)= q_0(s,\tau)\,\text{sech}\left[\Phi(s,\tau)\right]\exp\left[-i\theta(s,\tau)\right],  \label{fsol}
\end{equation}
with:
\begin{eqnarray}
 q_0(s,\tau)&=&\theta_0 G(s)H_i(\tau), \nonumber \\
 H_i(\tau)&=&\frac{\text{sech}^2\left(\Omega \tau \right)}{1+\Omega^2 \tanh^2\left(\Omega \tau \right)}, \label{fsola}
\end{eqnarray}

\begin{eqnarray}
\Phi(s,\tau)&=&16 L_0\int_0^{\tau}{H_r(t)H_i(t)dt} \nonumber \\
&-& 2H_i(\tau)F(s) - \ln\left[2H_i(\tau)\right], \nonumber \\
 H_r(\tau)&=&\frac{1}{\Omega^2}\frac{(1+\Omega^2)\tanh\left(\Omega \tau \right)}{1+\Omega^2\tanh^2\left(\Omega \tau \right)}, \label{fsolb}
\end{eqnarray}

\begin{eqnarray}
\theta(s,\tau)&=&8 L_0\int_0^{\tau}{\left[H_r^2(t) - H_i^2(t)\right]dt} \nonumber \\
&+& 2H_r(\tau)F(s). \label{fsolc}
\end{eqnarray}
We now examine profiles of the one-soliton solution obtained analytically in eq. (\ref{fsol}). To proceed we plot the analytical solution for arbitrary values of $L$ and $L_0$ defined in eqs. (\ref{inh}), with $\mu$ and $\mu_0$ expressed in formula (\ref{coef}) as functions of the thermo-optic coefficient $\beta$, the heat flux $Q$, the thermal conductivity $\kappa$, the material thickness $b$ and the input heat flux gradient $\gamma$. Values of these coefficients are reported in several recent experiments as we already indicated, except $\gamma$ for which we could not find experimental values in the literature. Nevertheless we shall assume that $\gamma$ can be nonzero, and treat the cases of positive and negative values of this parameter paying attention on the impact of this sign change on shape profiles of the one-soliton solution. It is Also useful to underline that because thermal lensing occurs along the $x$ axis, birefringence of the optical medium induced by thermal lensing is expected to affect beam propagation along this axis. Owing to the quadratic form of $V(x)$, birefringence of the optical medium will cause a double polarization of the beam along $x$ thus promoting bound states of two interacting pulses. To check this conjecture we plotted the field amplitude $A(x)=\sqrt{I(x)}$ as a function $x$, at selected positions along the transverse coordinate $z$. The integrals with respect to $\tau$ in eqs. (\ref{fsolb}) and (\ref{fsolc}) were computed numerically by means of a Simpson scheme, and we set the constant $\theta_0=1$ throughout simulations. \par
\begin{figure*}\centering
\begin{minipage}{0.52\textwidth}
\includegraphics[width=3.25in,height=2.2in]{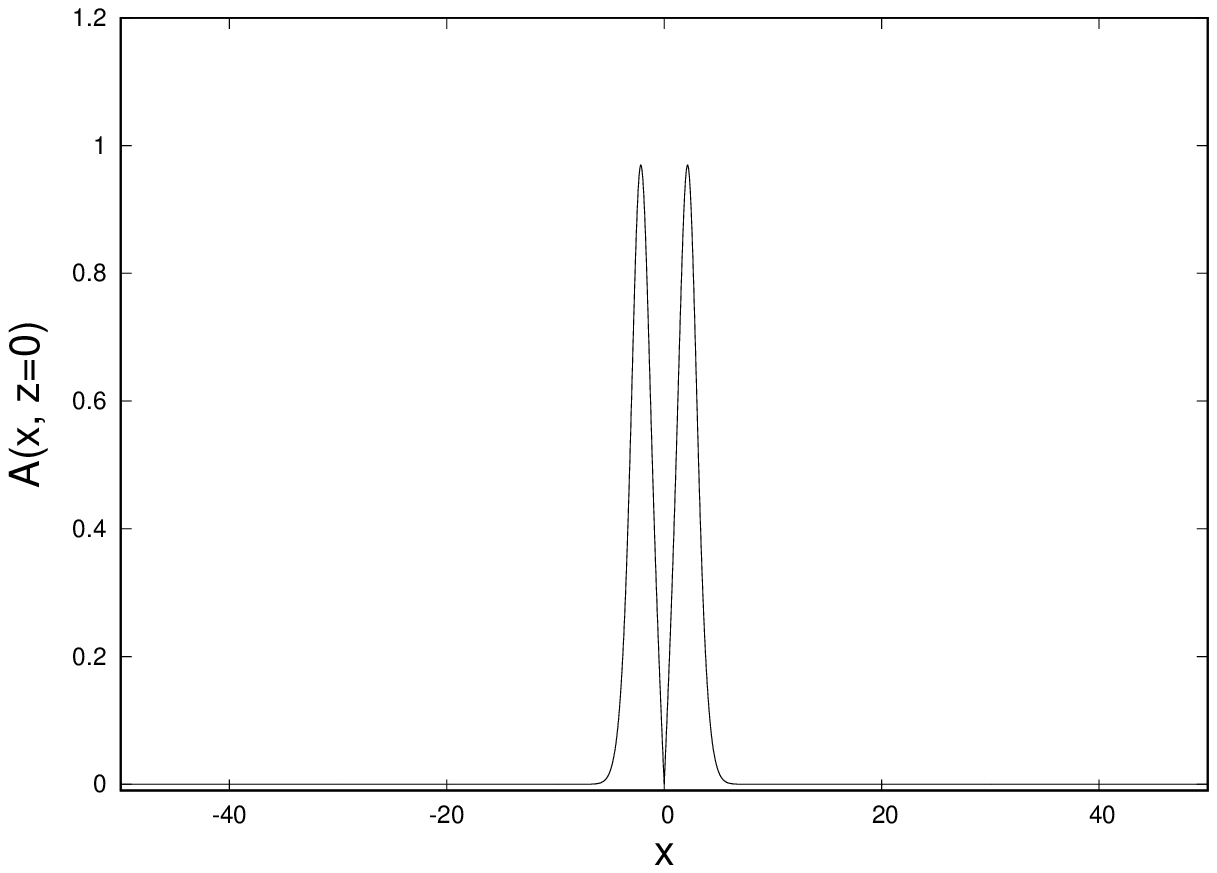}
\end{minipage}%
\begin{minipage}{0.52\textwidth}
\includegraphics[width=3.25in,height=2.2in]{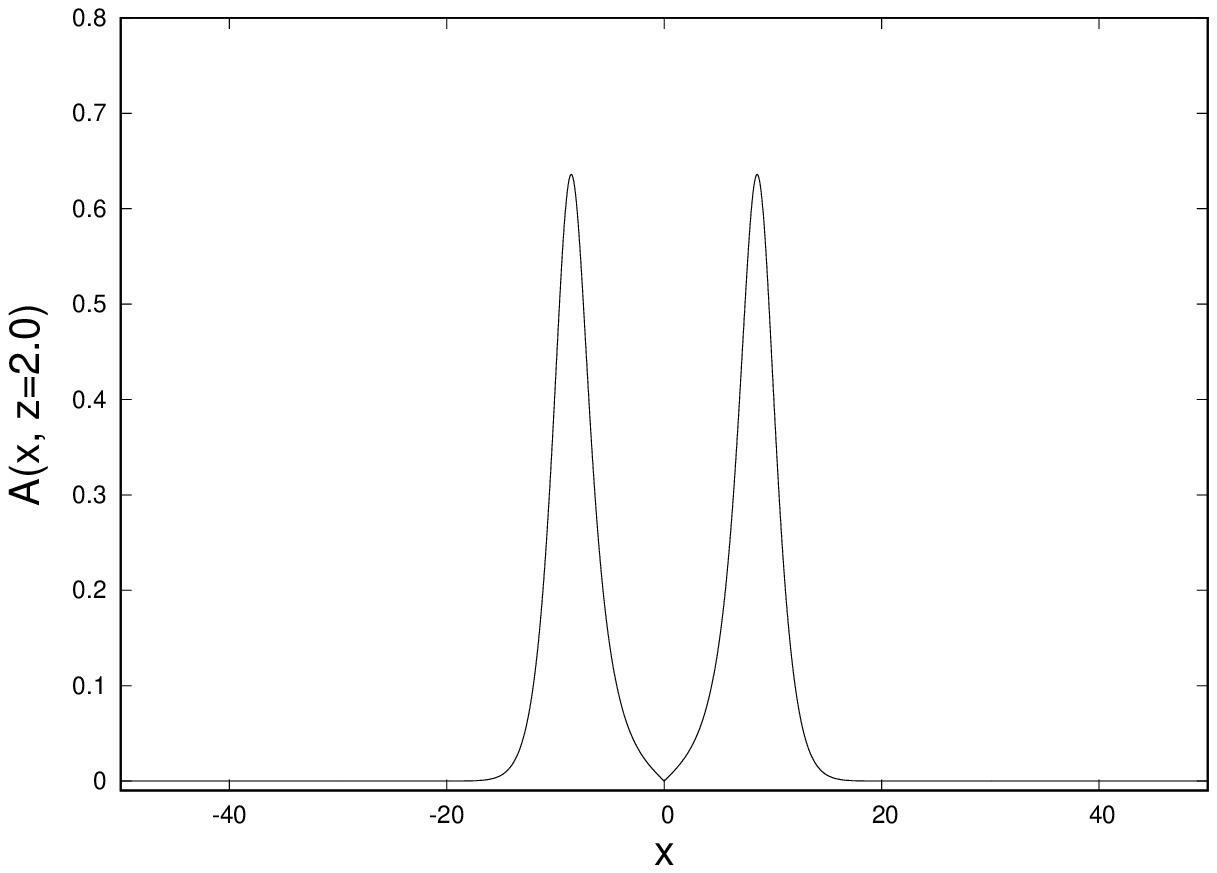}
\end{minipage}\\
\begin{minipage}{0.52\textwidth}
\includegraphics[width=3.25in,height=2.2in]{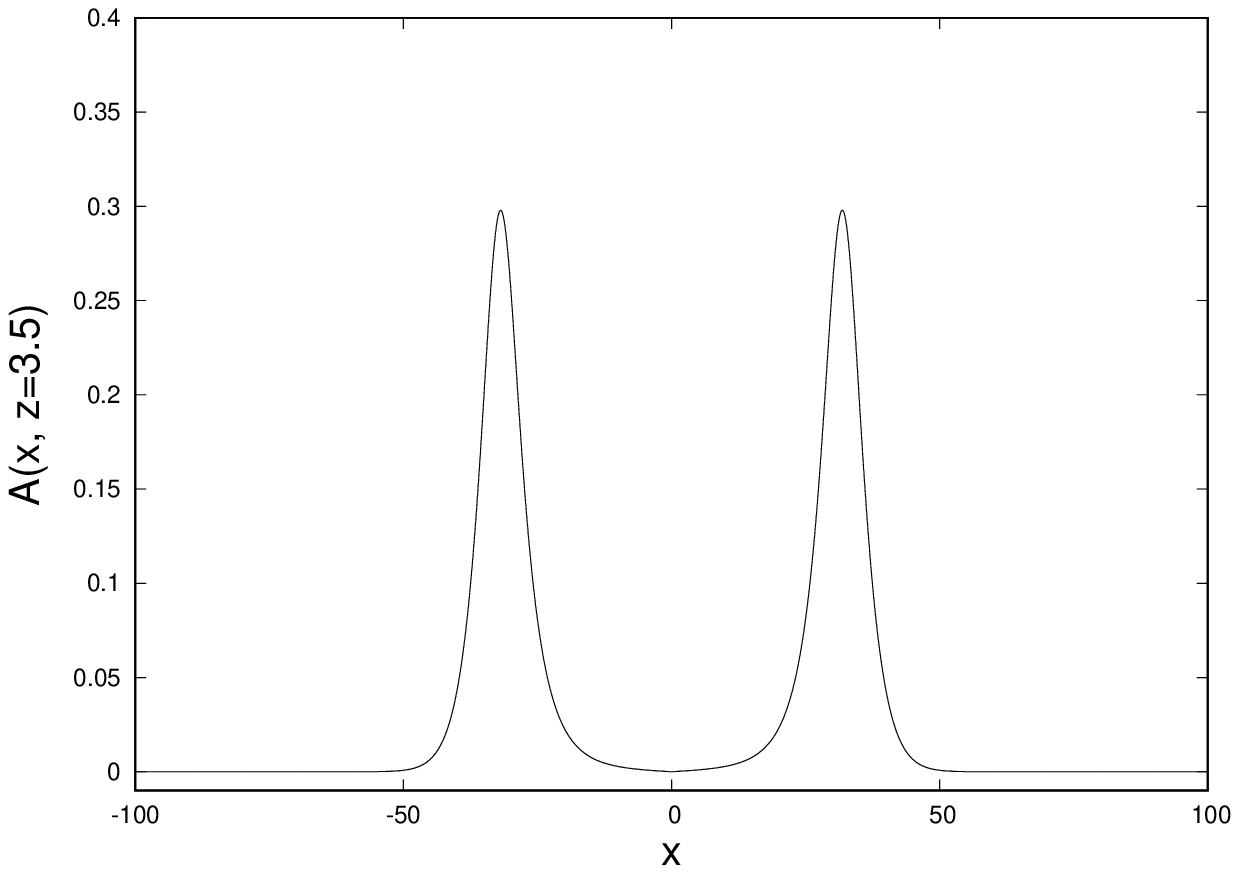}
\end{minipage}%  
\begin{minipage}{0.52\textwidth}
\includegraphics[width=3.25in,height=2.2in]{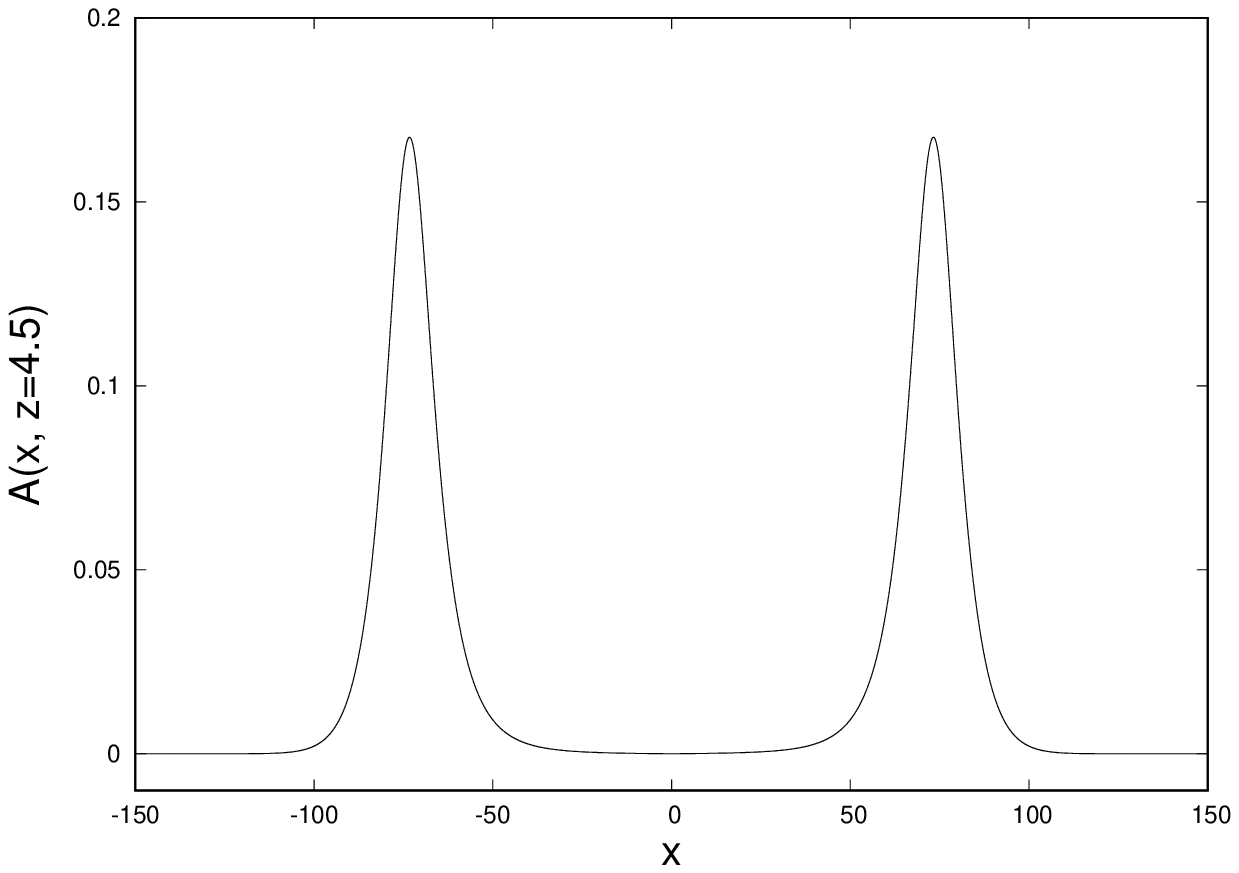}
\end{minipage}
\caption{\label{fig1}Shape of the one-soliton solution given by eq. (\ref{fsol}) as a function of $x$, for $\gamma=0$ and at different positions along the transverse axis $z$.}
\end{figure*}

Fig. \ref{fig1} shows graphs of the field amplitude $A(x)=\sqrt{I(x)}$ versus $x$ for $\gamma=0$, considering four different positions along the $z$ axis namely $z=0.0001$, 2, 3.5 and 4.5. Recall that for this figure the external potential $V(x)$ induced by thermal lensing effect is purely quadratic in $x$, and for our simulations we picked $L^2=0.5$.

Fig. (\ref{fig1}) features a two-pulse bound state representing a double-polarized structure or soliton molecule for the one-soliton solution eq. (\ref{fsol}), in agreement with our conjecture. On the figure we observe that when $z\approx 0$, amplitudes of the two pulses forming the soliton molecule overlap, resulting in a bound state of highly entangled twin pulses with zero separation between them. As the bound state moves away from the origin $z=0$, the twin pulses gradually separate and move aways from each other, with their coordinates $x$ always symmetric with respect to $x=0$. In the figure we clearly observe that as the entanglement of the bound pulse decays (i.e. as the separation between pulses increases), the tail of pulses decreases while their width at half tail increases. This is reminiscent of a relationship between similar parameters, for the one-soliton solution of the homogeneous nonlinear Schr\"odinger equation \cite{l2,re}.    
\par
\begin{figure*}\centering
\begin{minipage}{0.52\textwidth}
\includegraphics[width=3.25in,height=2.2in]{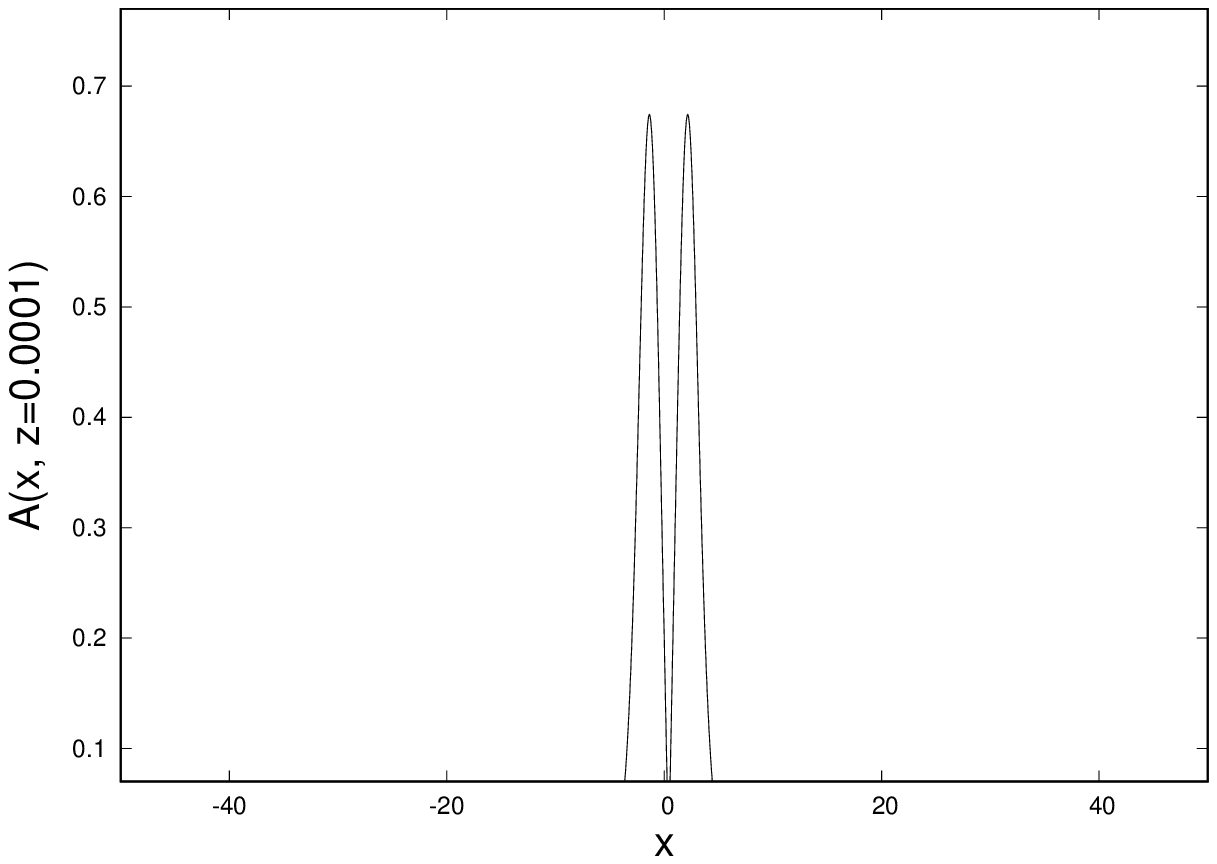}
\end{minipage}%
\begin{minipage}{0.52\textwidth}
\includegraphics[width=3.25in,height=2.2in]{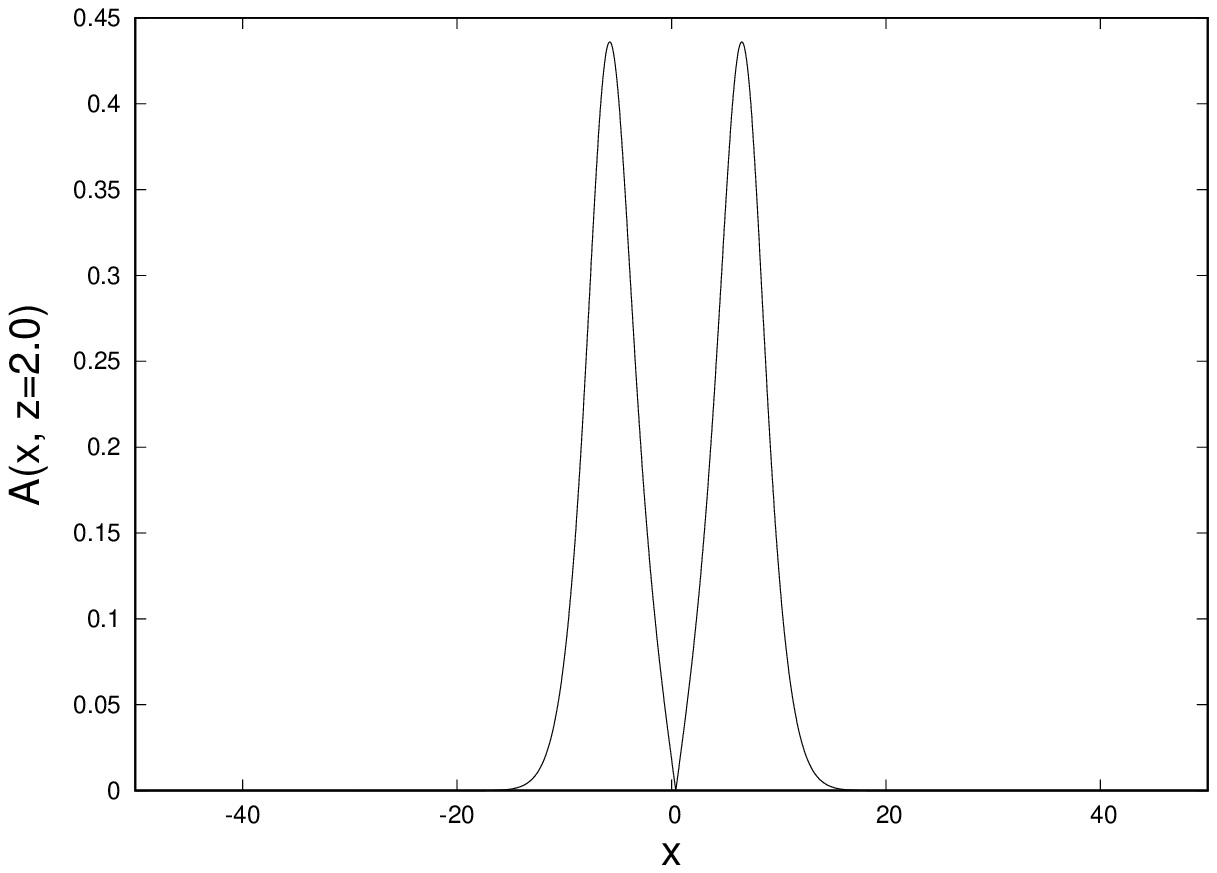}
\end{minipage}\\
\begin{minipage}{0.52\textwidth}
\includegraphics[width=3.25in,height=2.2in]{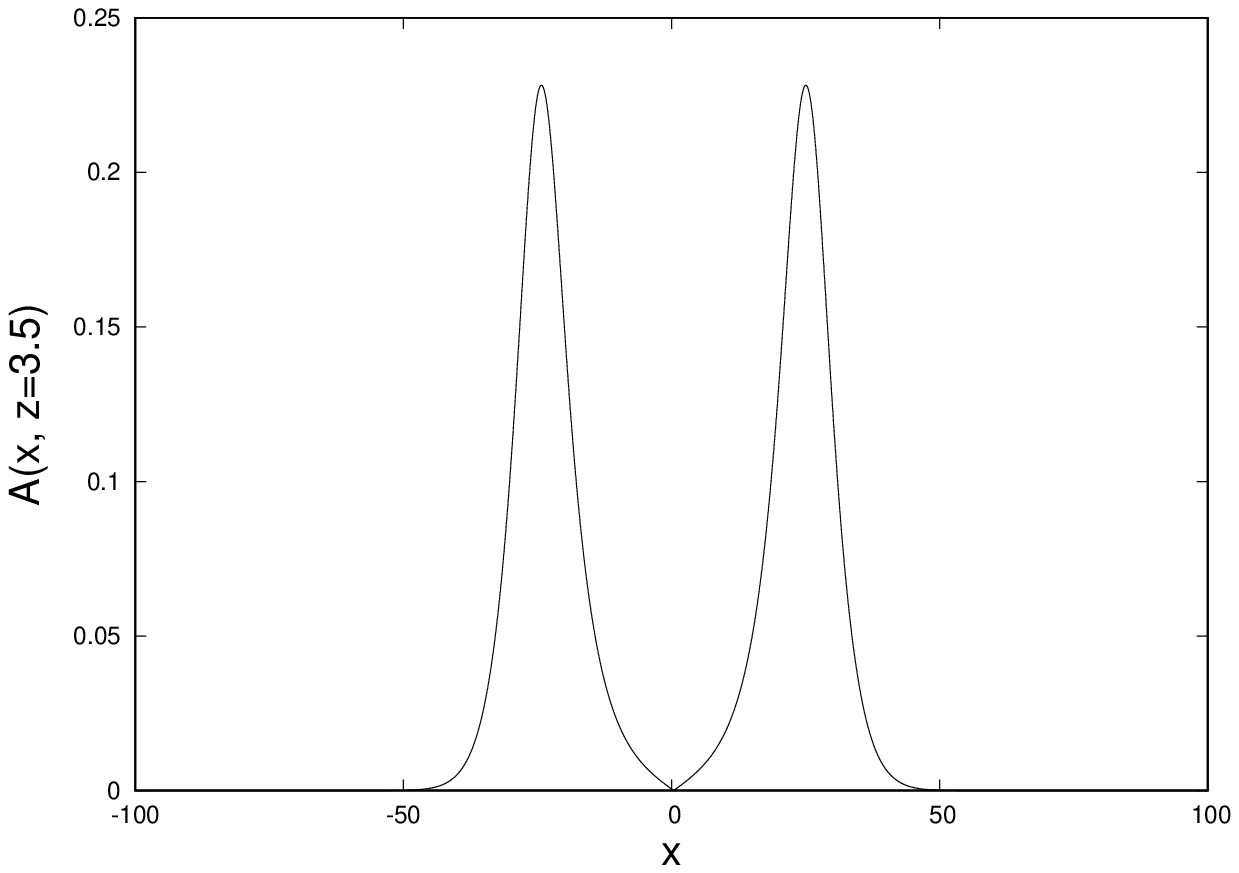}
\end{minipage}%   
\begin{minipage}{0.52\textwidth}
\includegraphics[width=3.25in,height=2.2in]{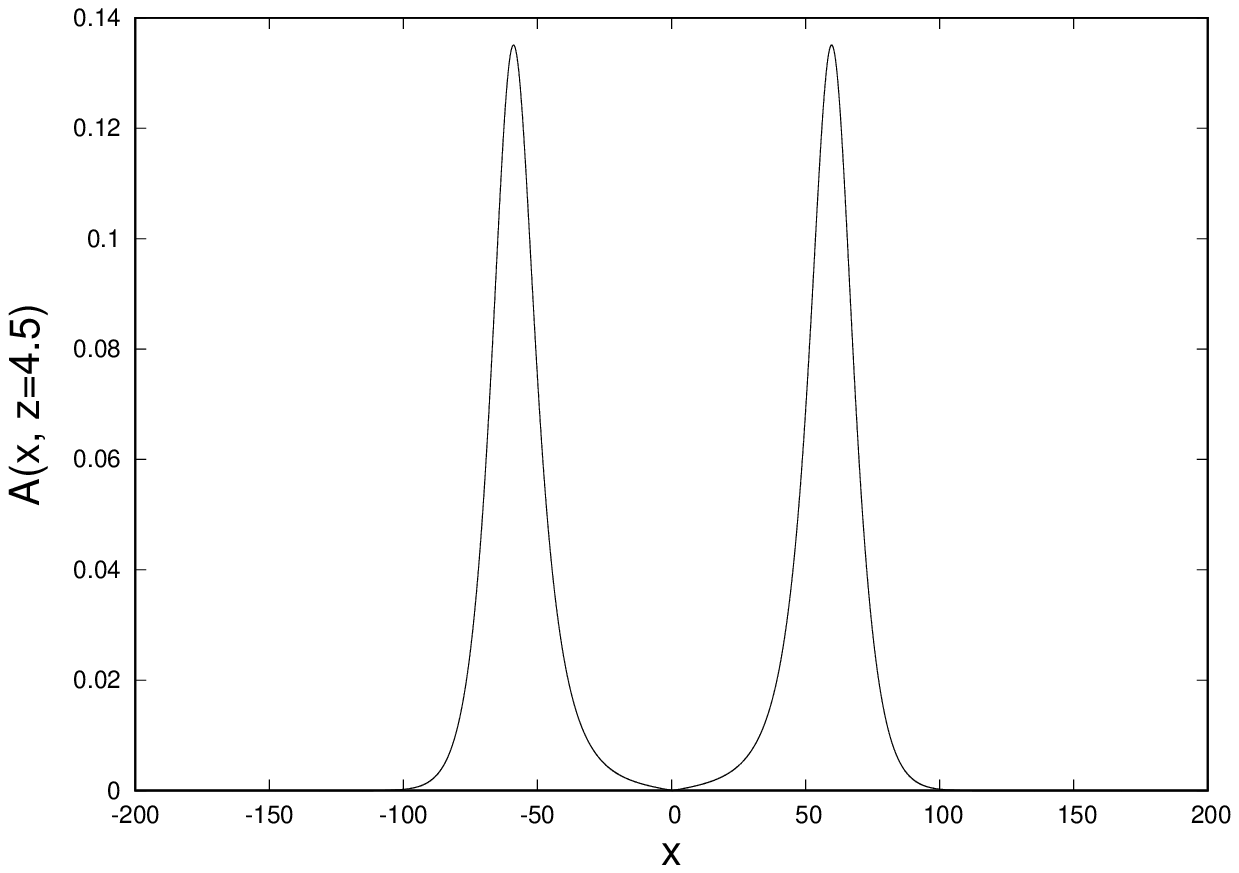}
\end{minipage}
\caption{\label{fig2} Shape of the one-soliton solution given by eq. (\ref{fsol}) as a function of $x$, for $\gamma=0.1$ and at different positions along the transverse axis $z$.}
\end{figure*}

\begin{figure*}\centering
\begin{minipage}{0.52\textwidth}
\includegraphics[width=3.25in,height=2.2in]{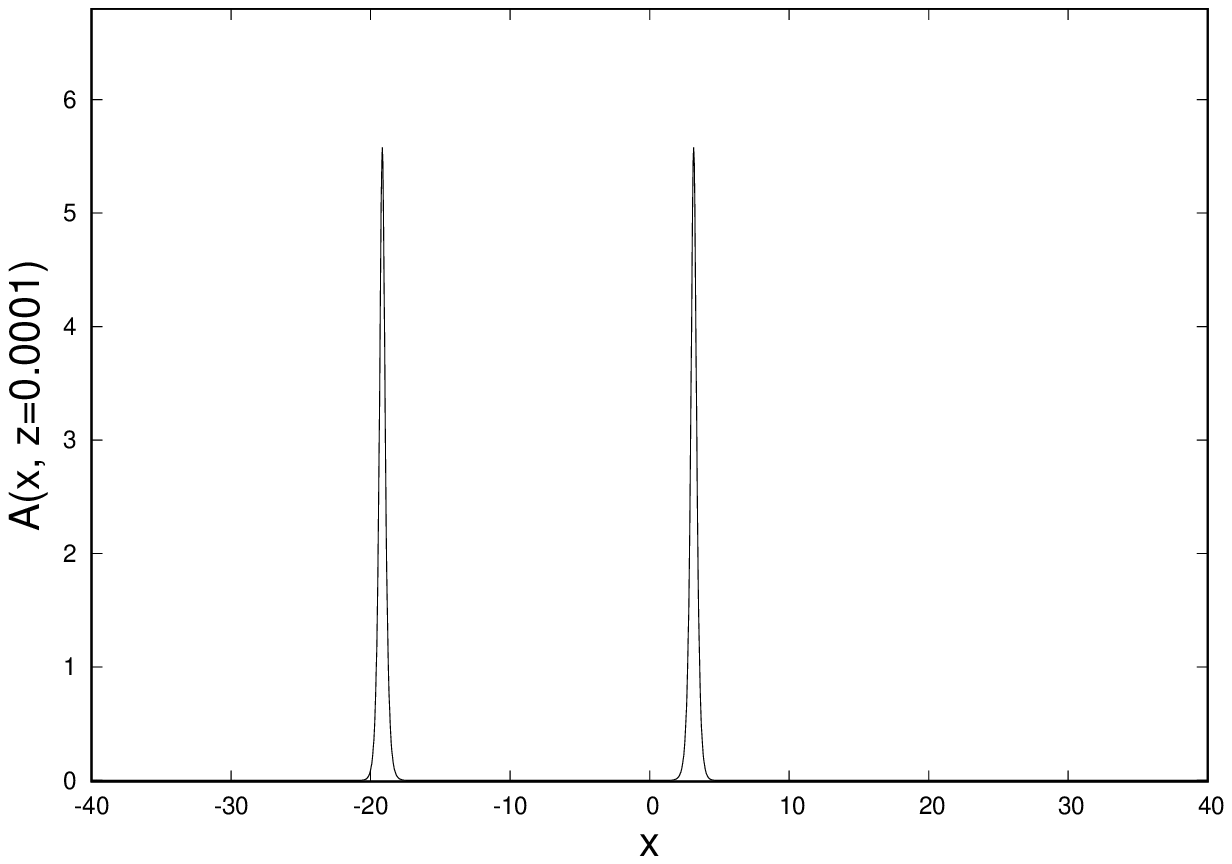}
\end{minipage}%
\begin{minipage}{0.52\textwidth}
\includegraphics[width=3.25in,height=2.2in]{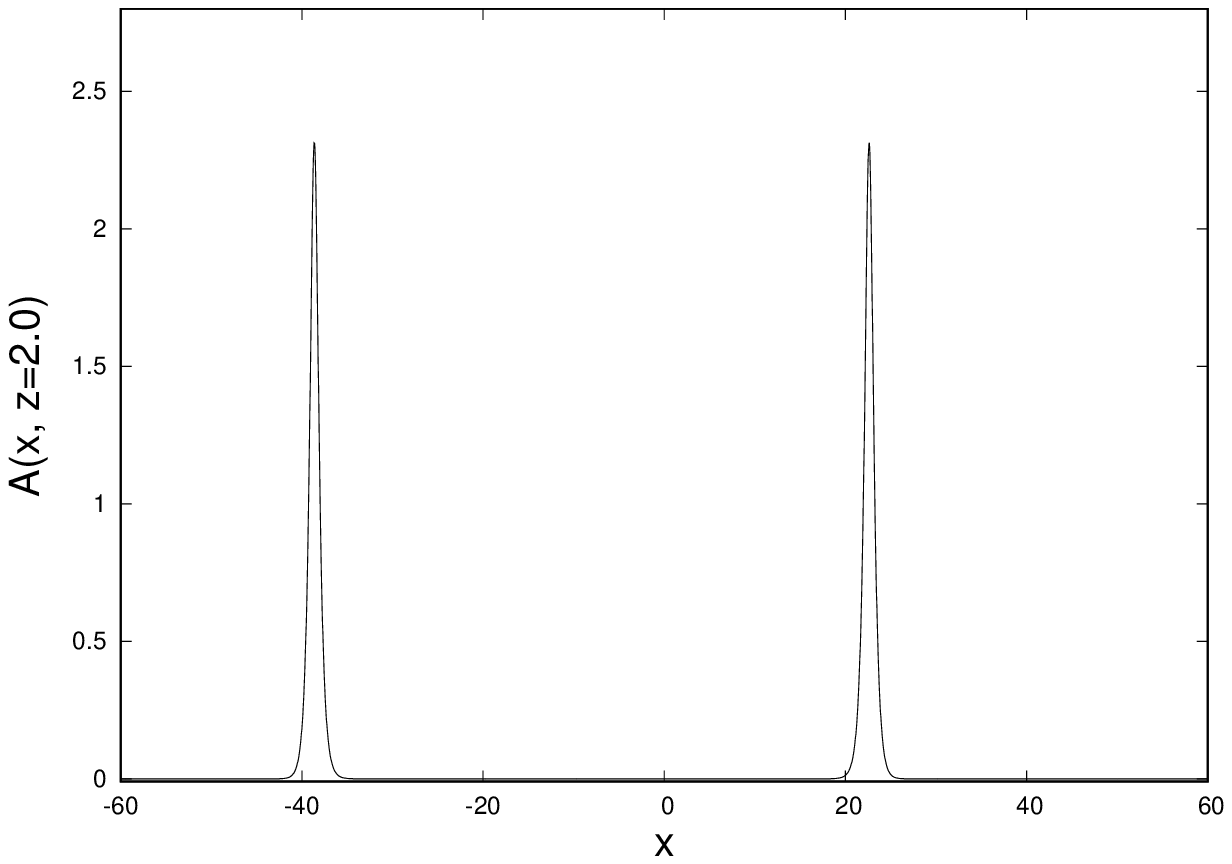}
\end{minipage}\\
\begin{minipage}{0.52\textwidth}
\includegraphics[width=3.25in,height=2.2in]{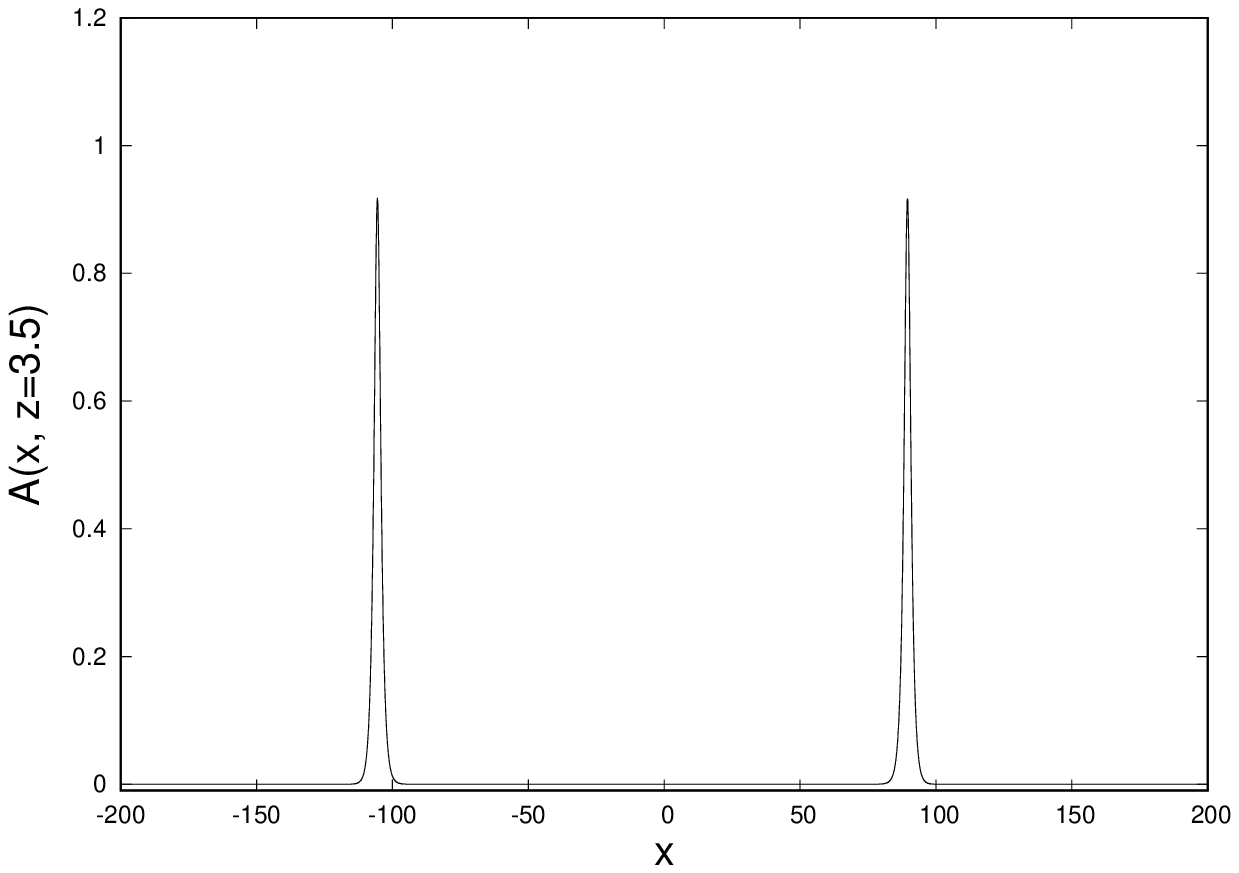}
\end{minipage}%
\begin{minipage}{0.52\textwidth}
\includegraphics[width=3.25in,height=2.2in]{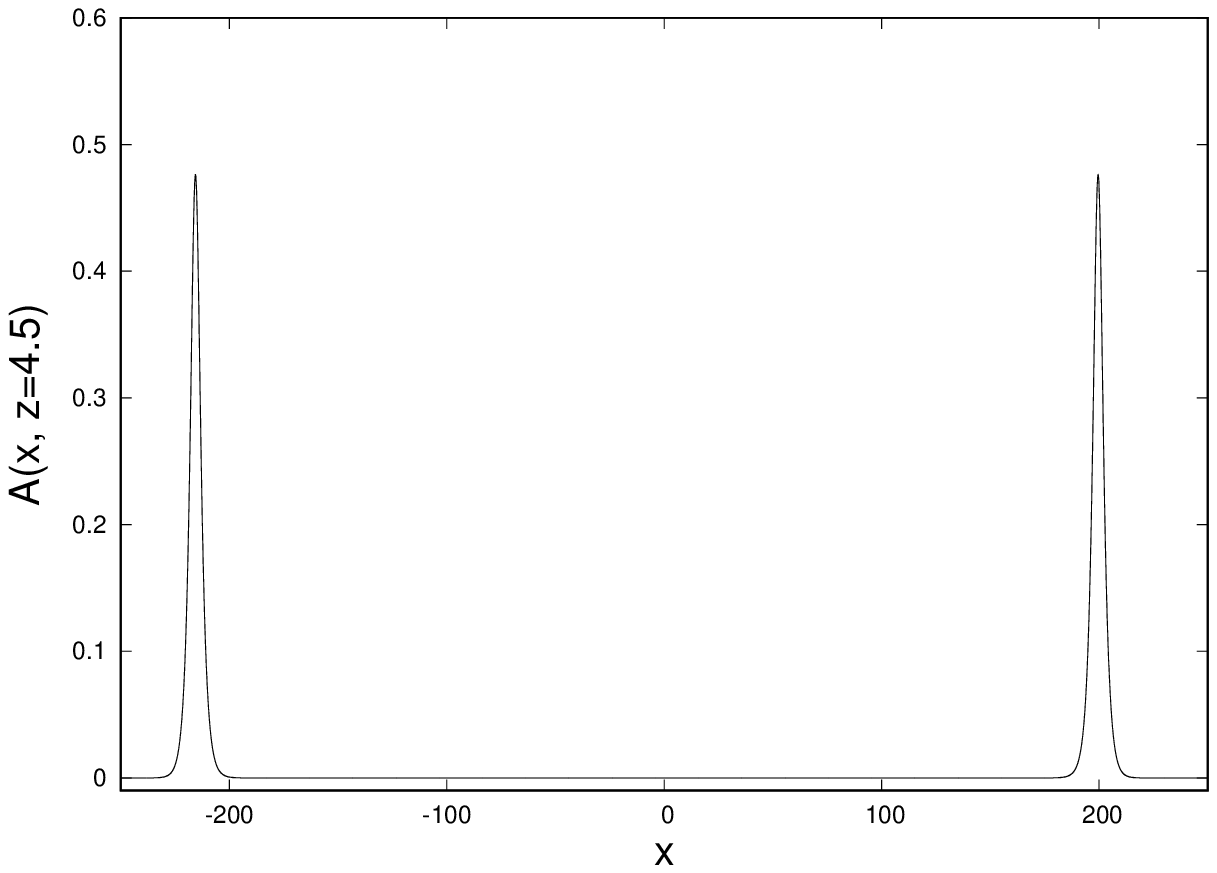}
\end{minipage}
\caption{\label{fig3} Shape of the one-soliton solution given by eq. (\ref{fsol}) as a function of $x$, for $\gamma=-2$ and at different positions along the transverse axis $z$. }
\end{figure*}

 To see the influence of a nonzero value of $\gamma$ on profiles of the soliton molecule, in fig. \ref{fig2} and fig. \ref{fig3} we plotted $A(x)=\sqrt{I(x)}$ as a function of $x$  now for $\gamma=0.1$ (fig. \ref{fig2}) and $\gamma=-2.0$ (fig. \ref{fig3}).   

As it clearly appears, a positive $\gamma$ compels the two pulses to propagate being always close to each other. So to say a positive value of $\gamma$ favors pulse entanglement in the soliton-molecule state, even for large values of $z$, compared with the situation observed in fig. \ref{fig1}. For graphs of fig. \ref{fig2} we have chosen $\gamma=0.1$, this of course is a small value but simulations show that large values of $\gamma$ lead to high entanglement of the pair of twin pulses, so high that the two pulses overlap. As opposed to this, when $\gamma$ is negative the two pulses are always well separated even close to the point $(x, z=0)$. Fig. \ref{fig3} shows that as $z$ increases the separation between pulses increases, with the two pulses now being asymmtrically positioned along the x axis. The pulse tails in this case change relatively slower than in the first case when $\gamma$ was zero, suggesting that this case is more suitable for propagation of the bound state with relatively moderate changes in pulse amplitudes. 
\par Our final comment in this section is devoted to a relevant feature of the analytical solution (\ref{fsol}), which emerged from its graphical representation in fig. \ref{fig3}, i.e. the asymmetric positions of pulses in the soliton bound state for negative $\gamma$. To understand this feature, recall that $\gamma$ was defined in such a way that the thermal-lensing induced external potential $V(x)$ is always symmetric in $x$ when $\gamma$ is zero, and asymmetric when $\gamma$ is nonzero. The symmetry of $V(x)$ implies that the probability of heat flow in the two directions along the axis of beam propagation is the same, whereas an asymmetric external potential $V(x)$ means that one direction will be priviledged. Choosing $\gamma$ negative in particular causes one component of the soliton molecule moving in the same direction with the heat flow, to be either "accelerated" or "slowed down" relatively with the other component moving opposite to the direction of heat flow. In fact these behaviours, and specifically the generation of symmetric or asymmetric soliton-molecule structures, are characteristic manifestations of thermo-optical phenomena occuring in the Kerr nonlinear material concomitantly with the propagation of heat-carrying high-intensity optical beams. \par 
From a more general standpoint, birefringence of optical media induced by thermal lensing is a quite common phenomenon in optoelectronic and photonic materials with sizable heat conductivity. This is for instance the case of optoelectronic devices, in which thermally induced birefringence has been shown to enable manipulations of polarization states of solid-state lasers \cite{ter1,ter2,ter3}. In the specific context of Nd:\ce{YVO4} self-mode-locked lasers, it was experimentally established \cite{ter1} that thermal lensing effects create an effective refractive index that varies quadrically with the laser focus position along the axis of propagation of the laser. Changes in optical indices of photonic waveguides, due to thermo-optical effects, have also been reported recently in several experimental as well as theoretical studies \cite{sil,phot1,phot2,phot3,phot4}. Thermal-lensing phenomena in optical media are therefore a subject of great current interest, their manifestations in $\beta$-\ce{Ga2O3} that emerged from the present study, offer new perspective for the design of photodetector solar cells extending up to high-power and ultrashort optical fields in the ultraviolet \cite{uvs}.            

\section{Conclusion}
\par In recent years there has been a great deal of interest in characteristic properties of $\beta$-\ce{Ga2O3}, an inorganic semiconducting isomorph of gallium oxide with outstanding thermo-optical properties. It is thus well established \cite{r12a} that $\beta$-\ce{Ga2O3} is a transparent material with a Kerr optical nonlinearity, on the other hand its thermo-optical properties, probed via measurements of thermo-optical effects caused by optical beam propagation along the axis of high thermal conductivity, point to a linear temperature dependence of its refractive index as suggested e.g. in ref. \cite{r12c}. Taking these experimental evidences into consideration, we proposed a theory aimed at predicting the appropriate shape profiles of high-intensity pulses that might propagate along the axis of heat flow in $\beta$-\ce{Ga2O3} material. Considering the governing Kerr nonlinearity and the quasi-one-dimensional structural anisotropy of the material, we described the field propagation by a cubic nonlinear Schr\"odinger equation in (1+1) dimensions \cite{nlq} where thermal lensing effects were represented by a space-dependent external potential. In view of the particular quadratic dependence of the external potential in beam position, we found that the resulting inhomogeneous cubic nonlinear Schr\"odinger equation ressembles an equation discussed by Balakrishnan \cite{ra}, for which the author developed a solution method based on the inverse-scattering transform approach. Exploiting this similarity, we obtained the analytcal expression of the exact one-soliton solution to our inhomogeneous cubic nonlinear Schr\"odinger equation. Using numerical simulations we established that this analytical solution features a soliton molecule originating from double-polarization of one sech-type pulse soliton upon propgation in the optical medium. We observed that by varying adequately characteristic parameters of the model, it was possible to control the degree of pulse entanglement as well as their widths and tails. 
\par It should be noted that in the present study, we focused on high-intensity optical fields with a bright-soliton feature. In some physical applications dark-soliton solutions, characterized by tanh-shaped profile, can be more appropriate. An analytical study similar to the present one, examining characteristic features of dark-soliton structures in the material, would enrich potential physical applications of $\beta$-\ce{Ga2O3} for integrated nonlinear optics \cite{r18}, photonics for solar technology involving high-power fields, and optoelectronics for high-power and high-repetition-rate (femtosecond and nanosecond) optical field processsing in the ultraviolet \cite{uvs}.

\section*{Conflict of interest}
The authors declare that they have no conflict of interest

\end{document}